\documentclass[pra,showpacs,amsfonts,amssymb,amsmath,superscriptaddress,twocolumn]{revtex4}

\usepackage{graphicx}
\usepackage{verbatim}

\newcommand{\ket}[1]{|{#1}\rangle}
\newcommand{\bra}[1]{\langle{#1}|}
\newcommand{\eq}[1]{(\ref{#1})}
\newcommand{\op}[1]{\hat{#1}}

\begin{document}

\title{Dynamics of dispersive single qubit read-out in circuit quantum electrodynamics}

\author{R.~Bianchetti}

\author{S.~Filipp}
\author{M.~Baur}
\author{J.~M.~Fink}
\author{M.~G\"oppl}
\author{P.~J.~Leek}
\author{L.~Steffen}
\affiliation{Department of Physics, ETH Zurich, CH-8093 Z{\"{u}}rich, Switzerland}
\author{A.~Blais}
\affiliation{D\'epartement de Physique, Universit\'e de Sherbrooke, J1K 2R1 Sherbrooke, Canada}
\author{A.~Wallraff}
\affiliation{Department of Physics, ETH Zurich, CH-8093 Z{\"{u}}rich, Switzerland}


\begin{abstract}

The quantum state of a superconducting qubit non-resonantly coupled to a transmission line resonator can be determined by measuring the quadrature amplitudes of an electromagnetic field transmitted through the resonator. We present experiments in which we analyze in detail the dynamics of the transmitted field as a function of the measurement frequency for both weak continuous and pulsed measurements. We find excellent agreement between our data and calculations based on a set of Bloch-type differential equations for the cavity field derived from the dispersive Jaynes-Cummings Hamiltonian including dissipation. We show that the measured system response can be used to construct a measurement operator from which the qubit population can be inferred accurately. Such a measurement operator can be used in tomographic methods to reconstruct single and multi qubit states in ensemble averaged measurements.
\end{abstract}

\maketitle

\section{Introduction} \label{Sec:intro}
Among several stringent requirements like scalability or precise coherent control, high-fidelity read-out of the qubit state is an important aspect of all experimental efforts in quantum information science~\cite{DiVincenzo2000}. For superconducting qubits \cite{Clarke2008} a number of read-out strategies~\cite{Vion2002, Martinis2002, Duty2004, Astafiev2004, Simmonds2004, Cooper2004, Sillanpaa2005, Siddiqi2006} specific to various implementations have been pursued. Early charge qubit read-outs implemented with single electron transistors (SET)~\cite{Devoret2000} were limited by strong current noise back-action from the measurement device on the qubit. Similar limitations applied to flux qubit read-out, measuring the switching current of a nearby superconducting quantum interference device (SQUID)~\cite{Wal2000, Chiorescu2003}. One strategy to achieve high read-out fidelities is to perform a quantum non-demolition (QND) measurement which preserves the eigenstates of the system Hamiltonian~\cite{Walls1994}. Repeated measurements yield identical results and consequently an improved signal-to-noise ratio. In the circuit quantum electrodynamics (QED) architecture, where a superconducting qubit is strongly coupled to a transmission line resonator~\cite{Blais2004,Wallraff2004b}, the qubit can both be controlled and read-out via the cavity using microwave signals. The read-out can be accomplished by detecting the dispersive qubit-state dependent shift of the resonator frequency~\cite{Wallraff2005}. In the dispersive limit, where the qubit transition frequency is far detuned from the resonator frequency, and for small photon numbers the measurement of the transmitted resonator field forms a QND measurement~\cite{Blais2004, Wallraff2004b, Wallraff2005, Gambetta2007, Gambetta2008}. Similarly, QND measurements have been employed to demonstrate explicitly the repeatability of this type of measurement for a flux qubit dispersively coupled to a nonlinear oscillator \cite{Lupascu2004, Lupascu2006, Lupascu2007}. In a related experiment the phase of a tank circuit coupled to a flux qubit was monitored, demonstrating resonant tunneling~\cite{Grajcar2004}. Note also, that in the circuit QED architecture a QND measurement has been proposed to generate and detect multi-qubit entangled states~\cite{Hutchison2009, Helmer2009, Bishop2009}.

Here we analyze the time-dependent response of the quadrature amplitudes of an electromagnetic field transmitted through a resonator to changes in the qubit state under dispersive interaction. We derive and discuss a set of Bloch-type equations describing accurately the dynamics of the qubit and the resonator for continuous measurements. We also extend the analysis to pulsed read-out which avoids measurement-induced dephasing during qubit manipulation and allows for stronger measurement. We analyze experimental data for different measurement frequencies and find excellent agreement with theory using a single set of independently measured parameters. This approach has already been successfully used in experiments \cite{Wallraff2005, Majer2007, Filipp2009b, Leek2009} but not yet discussed in literature.

Within this framework, we demonstrate our ability to infer the state of the qubit embedded in the cavity from a measurement of both field quadratures transmitted through the resonator. The construction of the corresponding projective measurement operator based on the state-dependent resonator response is outlined. For Rabi oscillation measurements, we discuss the extraction of the qubit state population and compare to numerical simulations.

\section{The physical system and its model} \label{Sec:Setup}

\begin{figure}[htb]
  \centering
    \begin{minipage}{1\linewidth}
    \includegraphics[width=0.95\textwidth]{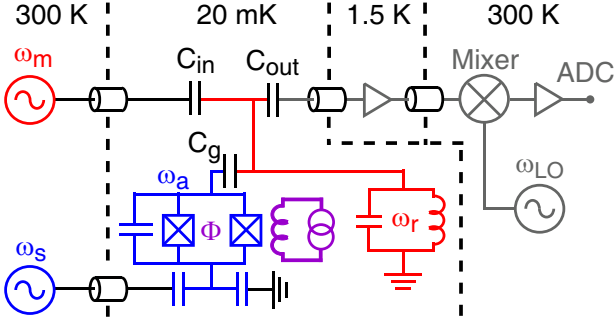}
    \end{minipage}
  \caption{(color online) Circuit diagram of the experimental setup. A harmonic oscillator modeled as an LC circuit with resonance frequency $\omega_{{\rm r}}$ is coupled to a transmon-type qubit~\cite{Koch2007} through the coupling capacitance $C_{{\rm g}}$. The qubit transition frequency $\omega_{{\rm a}}$ is controlled by an externally applied magnetic flux $\Phi$. The qubit state is coherently manipulated by a pulsed microwave source at the frequency $\omega_{{\rm s}}$. The resonator is probed by a signal applied to the input capacitor $C_{{\rm in}}$ at the frequency $\omega_{{\rm m}}$. The transmitted signal is amplified and downconverted by mixing with a local oscillator at frequency $\omega_{{\rm LO}}$ and then digitized using an analog to digital converter (ADC).}
    \label{fig:circuit}
\end{figure}

We consider a transmon-type qubit~\cite{Koch2007,Schreier2008} embedded in a transmission line resonator, as illustrated schematically in Fig.~\ref{fig:circuit}. The qubit is coupled to the resonator through the effective capacitance $C_{{\rm g}}$, leading to a qubit-resonator coupling of strength $g$ \cite{Wallraff2004b, Schoelkopf2008}. The qubit transition frequency $\omega_{{\rm a}}$ is tunable by an externally applied flux $\Phi$~\cite{Koch2007}. The resonator with resonance frequency $\omega_{{\rm r}}$ determined by its geometric and dielectric properties~\cite{Goppl2008} is modeled as an LC circuit.

When the qubit/resonator detuning $\Delta_{{\rm ar}} = \omega_{{\rm a}} - \omega_{{\rm r}}$ is much larger than the coupling strength $g$, this system is described by the dispersive approximation of the Jaynes-Cummings Hamiltonian~\cite{Blais2004}
\begin{equation} \label{eq:Disp}
H_{{\rm disp}} = \hbar (\omega_{{\rm r}} + \chi \hat{\sigma}_{{\rm z}} ) \hat{a}^\dagger \hat{a} + \frac{\hbar}{2}( \omega_{{\rm a}} + \chi ) \hat{\sigma}_{{\rm z}}.
\end{equation}
Here,
\begin{equation} \label{eq:chi}
\chi \approx - \frac{g^2 E_{{\rm c}}}{\Delta_{{\rm ar}} ( \Delta_{{\rm ar}} - E_{{\rm c}})}
\end{equation}
is the dispersive coupling strength between the resonator and the transmon qubit approximated as a two-level system~\cite{Koch2007}. $E_{{\rm c}}$ is the charging energy. The dispersive coupling leads to a qubit state-dependent shift of the resonator frequency, which we use to measure the qubit state.

As illustrated in Fig.~\ref{fig:circuit}, the qubit state is controlled by a coherent microwave field of amplitude $\Omega(t)$ and frequency $\omega_{{\rm s}}$ applied directly to the qubit while the measurement tone with amplitude $\epsilon_{{\rm m}} (t)$ and frequency $\omega_{{\rm m}}$ is applied to the input port of the resonator. These externally applied control fields are modeled by the Hamiltonian
\begin{equation} \label{eq:Control}
H_{{\rm d}} = \hbar \left( \epsilon_{{\rm m}} (t) \hat{a}^\dagger e^{-i \omega_{{\rm m}} t} + \Omega (t) \hat{\sigma}_+ e^{-i \omega_{{\rm s}} t} + \mathrm{h.c.} \right),
\end{equation}
where we have taken $\epsilon_{{\rm m}} (t)$ and $\Omega(t)$ to be real for simplicity of presentation.

The dynamics of the system in presence of dissipation and dephasing is described by a Lindblad-type master equation~\cite{Lindblad1976}
\begin{equation} \label{eq:Master}
\dot{\rho} = -\frac{i}{\hbar} [H,\rho] + \kappa \mathcal{D} [a] \rho + \gamma_1 \mathcal{D} [\hat{\sigma}_-] \rho + \frac{\gamma_\phi}{2} \mathcal{D} [\hat{\sigma}_{{\rm z}}] \rho \equiv \mathcal{L} \rho,
\end{equation}
where $H=H_{{\rm disp}}+H_{{\rm d}}$ and $\mathcal{D}[\hat{A}] \rho = \hat{A} \rho \hat{A}^\dagger - \hat{A}^\dagger \hat{A} \rho /2 - \rho \hat{A}^\dagger \hat{A} / 2$. Here, $\gamma_1 = 1 /T_1$ is the qubit decay rate,  $\gamma_\phi$ the qubit pure dephasing rate and $\kappa$ the photon decay rate. Since all experiments discussed in this paper are done at a small photon number $n \ll n_{{\rm ncrit}} = | \Delta_{{\rm ar}}^2 | /4g^2$ and as $\gamma_1$ exceeds the Purcell decay rate~\cite{Houck2008} we neglect higher-order corrections to this dispersive master equation~\cite{Boissonneault2008, Boissonneault2009}.

To study the dynamics of the coupled qubit/resonator system, we derive Bloch-like equations of motions for the expectation value of the qubit operators $\langle \hat{\sigma}_{{\rm i}} \rangle$ (${\rm i = x , y, z}$) and the resonator field operators $\langle \hat{a}\rangle$ and $\langle \hat{a}^\dagger \hat{a}\rangle$. However, the master equation~\eq{eq:Master} leads to an infinite set of coupled equations for these expectation values. For instance, the differential equation for $\langle \hat{a} \rangle$ involves terms proportional to $\langle \hat{a} \hat{\sigma}_{{\rm z}} \rangle$, $\langle \hat{a}^\dagger \hat{a} \hat{a} \hat{\sigma}_{{\rm z}} \rangle$ and $\langle \hat{a} \hat{\sigma}_{{\rm x}} \rangle$, which in turn involve even higher order terms. We therefore truncate this infinite series by factoring higher order terms $\langle \hat{a}^\dagger \hat{a} \hat{\sigma}_{{\rm i}} \rangle \approx \langle \hat{a}^\dagger \hat{a} \rangle\langle \hat{\sigma}_{{\rm i}} \rangle$ and $\langle \hat{a}^\dagger \hat{a} \hat{a} \hat{\sigma}_{{\rm i}} \rangle \approx \langle \hat{a}^\dagger \hat{a} \rangle\langle \hat{a} \hat{\sigma}_{{\rm i}} \rangle$, but keeping the terms $\langle \hat{a} \hat{\sigma}_{{\rm i}} \rangle$ which ensures that the field contains information about the qubit state. This choice of factorization yields the correct average values for coherent and Fock states~\cite{Boissonneault2007} and leads to a complete set of eight coupled differential equations
\begin{subequations}  \label{eq:cavity_bloch_n}
\begin{eqnarray}
d_t \langle \hat{a} \rangle & = & -i \Delta_{{\rm rm}} \langle \hat{a} \rangle - i \chi \langle \hat{a} \hat{\sigma}_{{\rm z}} \rangle - i \epsilon_{{\rm m}} - \frac{\kappa}{2} \langle \hat{a} \rangle, \label{eq:cavity_bloch_a} \\
d_t \langle \hat{\sigma}_{{\rm z}} \rangle & = & \Omega \langle \hat{\sigma}_{{\rm y}} \rangle - \gamma_1 \left( 1 + \langle \hat{\sigma}_{{\rm z}} \rangle \right), \label{eq:cavity_bloch_sz} \\
d_t \langle \hat{\sigma}_{{\rm x}} \rangle & = & - \left[ \Delta_{{\rm as}} + 2 \chi \left( \langle \hat{a}^\dagger \hat{a} \rangle + \frac{1}{2} \right) \right] \langle \hat{\sigma}_{{\rm y}} \rangle \nonumber \\
& & - \left( \frac{\gamma_1}{2} + \gamma_\phi \right) \langle \hat{\sigma}_{{\rm x}} \rangle, \label{eq:cavity_bloch_sx} \\
d_t \langle \hat{\sigma}_{{\rm y}} \rangle & = & \left[ \Delta_{{\rm as}} + 2 \chi \left( \langle \hat{a}^\dagger \hat{a} \rangle + \frac{1}{2} \right) \right] \langle \hat{\sigma}_{{\rm x}} \rangle \nonumber \\
& & -\left( \frac{\gamma_1}{2} + \gamma_\phi \right) \langle \hat{\sigma}_{{\rm y}} \rangle - \Omega \langle \hat{\sigma}_{{\rm z}} \rangle, \label{eq:cavity_bloch_sy} \\
d_t \langle \hat{a} \hat{\sigma}_{{\rm z}} \rangle & = & - i \Delta_{{\rm rm}} \langle \hat{a} \hat{\sigma}_{{\rm z}} \rangle - i \chi \langle \hat{a} \rangle + \Omega \langle \hat{a} \hat{\sigma}_{{\rm y}} \rangle \nonumber \\
& & -i \epsilon_{{\rm m}} \langle \hat{\sigma}_{{\rm z}} \rangle - \gamma_1 \langle \hat{a} \rangle - \left( \gamma_1 + \frac{\kappa}{2} \right) \langle \hat{a} \hat{\sigma}_{{\rm z}} \rangle, \label{eq:cavity_bloch_asz} \\
d_t \langle \hat{a} \hat{\sigma}_{{\rm x}} \rangle & = & - i \Delta_{{\rm rm}} \langle \hat{a} \hat{\sigma}_{{\rm x}} \rangle - \left[ \Delta_{{\rm as}} + 2 \chi \left( \langle \hat{a}^\dagger \hat{a} \rangle + 1 \right) \right] \langle \hat{a} \hat{\sigma}_{{\rm y}} \rangle \nonumber \\
& & - i \epsilon_{{\rm m}} \langle \hat{\sigma}_{{\rm x}} \rangle - \left( \frac{\gamma_1}{2} + \gamma_{\phi} + \frac{\kappa}{2} \right) \langle \hat{a} \hat{\sigma}_{{\rm x}} \rangle, \label{eq:cavity_bloch_asx} \\
d_t \langle \hat{a} \hat{\sigma}_{{\rm y}} \rangle & = & - i \Delta_{{\rm rm}} \langle \hat{a} \hat{\sigma}_{{\rm y}} \rangle + \left[ \Delta_{{\rm as}} + 2\chi \left( \langle \hat{a}^\dagger \hat{a} \rangle + 1 \right) \right] \langle \hat{a} \hat{\sigma}_{{\rm x}} \rangle \nonumber \\
& & - i \epsilon_{{\rm m}} \langle \hat{\sigma}_{{\rm y}} \rangle  - \left( \frac{\gamma_1}{2} + \gamma_{\phi} + \frac{\kappa}{2} \right) \langle \hat{a} \hat{\sigma}_{{\rm y}} \rangle \nonumber \\
& & - \Omega \langle \hat{a} \hat{\sigma}_{{\rm z}} \rangle, \label{eq:cavity_bloch_asy} \\
d_t \langle \hat{a}^\dagger \hat{a} \rangle & = & - 2 \epsilon_{{\rm m}} {\rm Im}\langle \hat{a} \rangle - \kappa \langle \hat{a}^\dagger \hat{a} \rangle ,
\end{eqnarray}
\end{subequations}
which we refer to as \emph{Cavity-Bloch equations}. Here, we have defined $\Delta_{{\rm as}} = \omega_{{\rm a}}-\omega_{{\rm s}}$ and $\Delta_{{\rm rm}} = \omega_{{\rm r}}-\omega_{{\rm m}}$ as the detuning of the control and measurement microwave fields from the qubit and cavity frequency, respectively. While these equations are apparently more complex than Eq.~(\ref{eq:Master}), they can be analytically solved in some cases and are much faster to solve numerically. Note, that they do not include measurement-induced dephasing caused by photon shot-noise~\cite{Blais2004,Gambetta2006}, because only the expectation value of $\hat{a}^\dagger \hat{a}$ is taken into account, and higher order moments are omitted. This is of no consequence for the understanding of the experiments presented here because of the small photon numbers present during the measurement. In experiments where measurement-induced dephasing is important, Eq.~\eq{eq:Master} has to be solved directly~\cite{Gambetta2008}, or higher order terms have to be taken into account~\cite{Boissonneault2008}.

\section{Continuous measurement response} \label{Sec:ContMeas}
\begin{figure}[b]
\centering
    \begin{minipage}{1\linewidth}
    \includegraphics[width=1\textwidth]{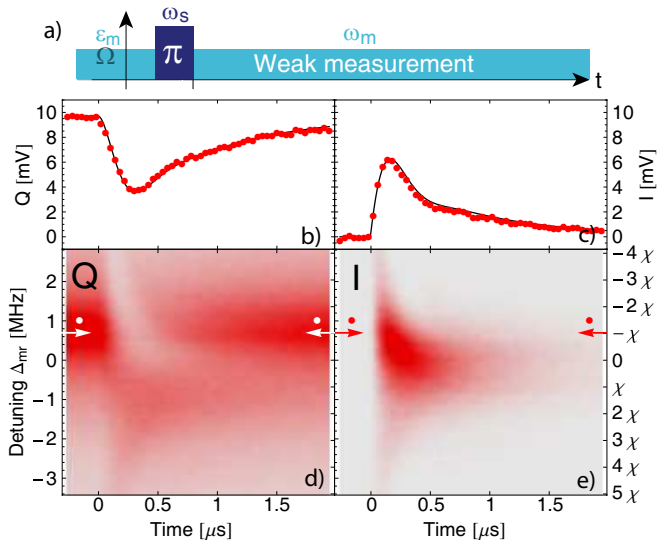}
        \end{minipage}
        \caption{(color online) a) In a continuous measurement, the cavity is driven with the qubit in its ground state, populating the resonator with $\bar{n}=1$ photons on average at the cavity resonance frequency. The qubit is then prepared in the excited state with a $\pi$-pulse ending at $t=0$. b) and c) averaged measurement response $Q$, $I$ versus time $t$ for a continuous weak measurement at the frequency $\omega_{{\rm m}} = \omega_{{\rm r}} - \chi$.  Solid lines show the predicted response from the Cavity-Bloch equations, Eq.~(\ref{eq:cavity_bloch_n}). Time resolved data taken at different detunings $\Delta_{{\rm mr}}$ is shown in d) and e). The arrows indicate the detuning at which the data shown in b) and c) is taken. The colormap codes red for a positive amplitude and white for zero.}
    \label{fig:cont_meas_detuning}
\end{figure}
\begin{figure}[b]
  \centering
    \begin{minipage}{1\linewidth}
    \includegraphics[width=0.95\textwidth]{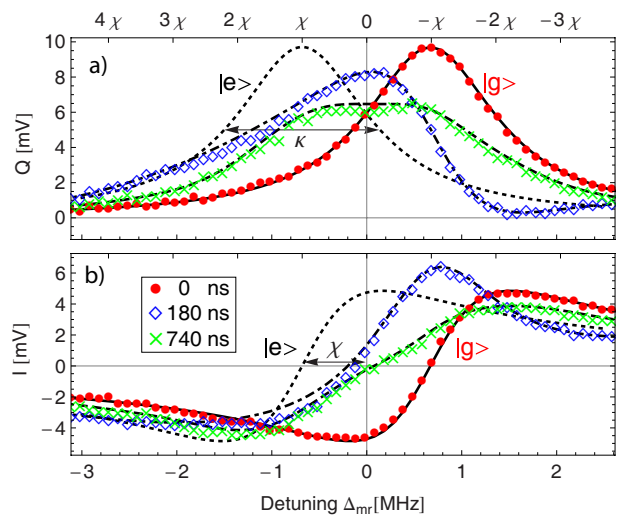}
    \end{minipage}
  \caption{(color online) Transmission spectrum of the resonator. The $Q$ quadrature of the field is shown in a) and the $I$ quadrature in b). The datapoints represent the instantaneous averaged response of the field before (red solid points), 180~ns (blue diamonds) and 740~ns after (green crosses) the $\pi$-pulse is applied. The underlying lines show the numerical simulations and the dotted line shows the expected response of the system for the qubit in the excited state $\ket{e}$ with infinite lifetime.}
    \label{fig:cavity_pull}
\end{figure}

To experimentally determine the state of the qubit, we probe the dynamics of the resonator-qubit system by measuring the resonator transmission at different frequencies $\omega_{{\rm m}}$. A time resolved, phase sensitive measurement of the transmission quadrature amplitudes is realized by down converting the measurement signal at frequency $\omega_{{\rm m}}$ in a mixer to an intermediate frequency $\Delta_{{\rm mLO}} = \omega_{{\rm m}} - \omega_{{\rm LO}} = 25~{\rm MHz}$ using a local oscillator of frequency $\omega_{{\rm LO}}$ and phase $\phi$, yielding one independent data point every $40\,\rm{ns}$, see Fig~\ref{fig:circuit}. This intermediate frequency (IF) signal is digitized in an analog-to-digital converter (ADC). The whole experiment is repeated and averaged 650'000 times to enhance the signal-to-noise-ratio and then digitally down converted to DC, leading to a measurement time of 26 ms per point for a total of 6.5 seconds for 250 points in the trace. From this we obtain both the in-phase ($I$) and the quadrature ($Q$) components of the transmitted field $A \sin ( \Delta_{{\rm mLO}} t + \phi) \equiv I \sin \Delta_{{\rm mLO}} t + Q \cos \Delta_{{\rm mLO}} t$. Using input-output theory~\cite{Walls1994}, these quadratures at the output of the resonator are related to the cavity Bloch equations by
\begin{eqnarray} \label{eq:IQdef}
I (t) & = & \sqrt{ Z \hbar \omega_{{\rm r}} \kappa} \ {\rm Re} \langle \hat{a}(t) \rangle, \nonumber \\
Q (t) & = & \sqrt{ Z \hbar \omega_{{\rm r}} \kappa} \ {\rm Im} \langle \hat{a}(t) \rangle ,
\end{eqnarray}
where $Z$ is the characteristic impedance of the transmission line connected to the resonator.

When arbitrary qubit rotations can be performed, it is sufficient to consider the measurement response for the qubit prepared in either its ground $|g\rangle$ or excited state $|e \rangle$ for a full characterization of the qubit state~\cite{James2001}. Figure~\ref{fig:cont_meas_detuning}a) shows the pulse scheme used for the measurement. The time dependent quadrature amplitudes $I$ and $Q$ are measured at the cavity resonance frequency with the qubit in the ground state ($\omega_{{\rm m}} = \omega_{{\rm r}} - \chi$). The resonator is continuously driven at a measurement drive amplitude of $\epsilon_{{\rm m}}^2 = \kappa / 2$, populating the resonator with $\bar{n} \approx 1$ photons on average in resonance. A $10~{\rm ns}$ long $\pi$ pulse ending at time $t=0$ and resonant with the ac-Stark \cite{Schuster2005} and Lamb shifted \cite{Fragner2008} qubit transition frequency $\omega_{{\rm s}} = [\omega_{{\rm a}} + 2\chi ( \langle a^\dag a\rangle + 1/2) ]:=\omega_{{\rm s,res}}$ is then applied to the qubit, see Figs.~\ref{fig:cont_meas_detuning}b) and c). Qubit relaxation during the $\pi$ pulse limits the achievable $|e \rangle$ state population to 99\%~\cite{Chow2009}, as obtained by solving the Cavity-Bloch equations. This is within the statistical uncertainty of the detection. Furthermore, thermal excitations of the qubit are expected to be very low and are therefore neglected.

The dependence of the quadrature components $I$ and $Q$ on the detuning $\Delta_{{\rm mr}}$ of the measurement frequency from the bare resonator frequency is plotted in Figs.~\ref{fig:cont_meas_detuning}d) and e). For clarity, the quadratures are rotated in the $IQ$-plane for each measurement frequency $\omega_{{\rm m}}$ such that the $Q$ quadrature is maximal in the steady-state (qubit in the ground state), resulting in $Q=A$ and $I=0$ for $t \rightarrow \infty$. As a result, before the $\pi$-pulse the $I$ quadrature is always zero.

The time and frequency dependence of the measurement signal is accurately described by the Cavity-Bloch equations with a single set of independently measured, non adjustable parameters as indicated by the solid lines in Figs.~\ref{fig:cont_meas_detuning}b) and c). The cavity resonance frequency is determined as $\omega_{{\rm r}}/ 2 \pi = 6.44252\pm 0.00002~{\rm GHz}$ with a photon decay rate of $\kappa / 2 \pi = 1.69 \pm 0.02~{\rm MHz}$. The qubit transition frequency is determined spectroscopically as $\omega_{{\rm a}} / 2 \pi \approx 4.009 \pm 0.001~{\rm GHz}$ with a charging energy of $E_{{\rm c}} / h = 232.5 \pm 0.5~{\rm MHz}$~\cite{Schreier2008}. The transition frequency is adjusted using external magnetic flux. The qubit-cavity coupling $g / 2 \pi = 134 \pm 1~{\rm MHz}$ is extracted from a measurement of the vacuum-Rabi mode splitting at $\omega_{{\rm a}} = \omega_{{\rm r}}$~\cite{Wallraff2004b}.

The cavity pull $\chi / 2 \pi = -0.69 \pm 0.02~{\rm MHz}$ is determined spectroscopically. This is done by measuring the cavity resonance frequency leaving the qubit in the ground state and then measuring its frequency shift applying a continuous coherent tone to the effective qubit transition frequency $\omega_{{\rm s,res}}$. When the qubit transition is saturated ($\Omega \gg \gamma_1$), the resonator is shifted on average by $\chi$. This value is in good agreement with the full transmon model taking into account higher levels~\cite{Koch2007} ($\chi / 2 \pi = -0.71~{\rm MHz}$).

In fitting the measurement response in Fig.~\ref{fig:cont_meas_detuning}, the qubit decay rate $\gamma_1 / 2 \pi=0.19\pm 0.01~{\rm MHz}$ is used as an adjustable parameter which is equal to an independent measurement of $\gamma_1$ within the statistical uncertainty. In practice, the qubit decay rate is determined for one measured trace, and then kept fixed for all other traces. Note that, for short $\pi$-pulses, the dephasing rate $\gamma_\phi$ has no measurable influence on the solution of the equations. Additionally, a single scaling factor is introduced to relate the quadrature voltages at the output of the resonator to the digitized voltages after amplification.

To interpret the time and frequency dependence of the measurement signal shown in Fig.~\ref{fig:cont_meas_detuning} it is instructive to plot $I$ and $Q$ as a function of $\omega_{{\rm m}}$ at fixed times $t$, as shown in Fig.~\ref{fig:cavity_pull}. With the qubit in $\ket{g}$, red points in Fig.~\ref{fig:cavity_pull}, the resonator transmission exhibits the expected line shapes for both quadratures. When applying a $\pi$-pulse to prepare the qubit in $\ket{e}$, the cavity resonance frequency shifts by $2\chi$, but the transmitted quadratures respond only on a time scale corresponding to the photon lifetime $T_{\kappa} \equiv 1/\kappa$. The lineshape of the cavity transmission spectrum centered at $+\chi$ will only be reached in the limit of $T_{1} \gg T_{\kappa}$, see dotted line in Fig.~\ref{fig:cavity_pull}. The interplay of the cavity field rise time and the qubit decay time results in the observed dynamics of the cavity transmission in Figs.~\ref{fig:cont_meas_detuning} and~\ref{fig:cavity_pull}.

At time $t=180~\rm{ns} \sim 1.9 \, T_{\kappa} \sim 0.2 \, T_1$ after the preparation of $\ket{e}$ the shift of the cavity resonance to lower frequency towards $+\chi$ is clearly visible, see blue diamonds in Fig.~\ref{fig:cavity_pull}. At $t=740~\rm{ns} \sim 7.9 \, T_{\kappa} \sim 0.9 \, T_1$, when $\approx 60\%$ of the excited state qubit population  $P_e$ is decayed, the measured curve is approximately the average between the steady state $\ket{g}$ and $\ket{e}$ responses, see green crosses in Fig.~\ref{fig:cavity_pull}.

When looking at the time traces in Fig.~\ref{fig:cont_meas_detuning}d), the effective shift of the resonance to lower frequency explains the reduction of the signal in the $Q$ quadrature for measurement detunings $\Delta_{{\rm mr}} >  -0.6~\rm{MHz} \sim \chi $. For $\Delta_{{\rm mr}} < -0.6~\rm{MHz}$ the amplitude is increased after the $\pi$-pulse because the resonator is driven closer to resonance. Given our choice of the rotation of the traces in the $IQ$-plane, the $I$ quadrature of Fig.~\ref{fig:cont_meas_detuning}e) acts like a phase and always shows a positive response to the $\pi$-pulse.

The same considerations explain the features seen in the single measurement trace in Figs.~\ref{fig:cont_meas_detuning}b) and c) taken at a measurement frequency corresponding to $\Delta_{{\rm mr}} = -\chi$. The change of the $I$ and $Q$ quadratures on a timescale $T_{\kappa}$ after the $\pi$-pulse reflects the relaxation of the field in response to the qubit excitation. The time scale of the return of the quadratures to their initial values is determined by the qubit decay at rate $\gamma_1$.

\section{Pulsed measurement response} \label{Sec:Pulsed_Measurements}
\begin{figure}[t]
    \centering
    \begin{minipage}{1\linewidth}
        \includegraphics[width=0.9\textwidth]{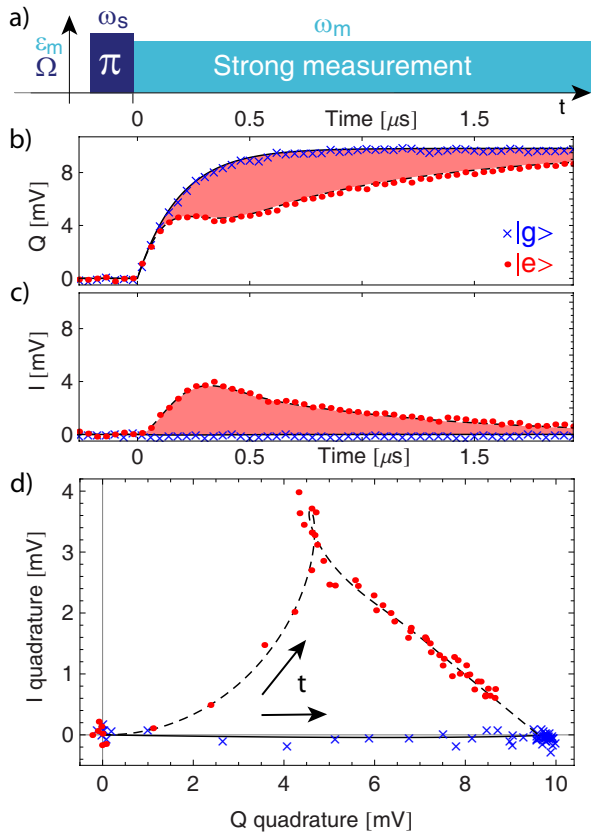}
    \end{minipage}
    \caption{(color online) a) Pulsed measurement scheme. b),c) Measurement response versus time for a pulsed averaged measurement taken in the same conditions as Fig.~\ref{fig:cont_meas_detuning} for the $Q$ and $I$ quadrature respectively. Red points show the response for the qubit being prepared in the excited state $| e \rangle$. The blue crosses are the measured response to the qubit prepared in $| g \rangle$. The trajectory of the field in the $IQ$-plane is plotted in d), the arrows indicate the direction of the time.}
  \label{fig:pulsed_iqtrace}
\end{figure}
\begin{figure*}[t]
    \centering
    \begin{minipage}{1\linewidth}
        \includegraphics[width=1\textwidth]{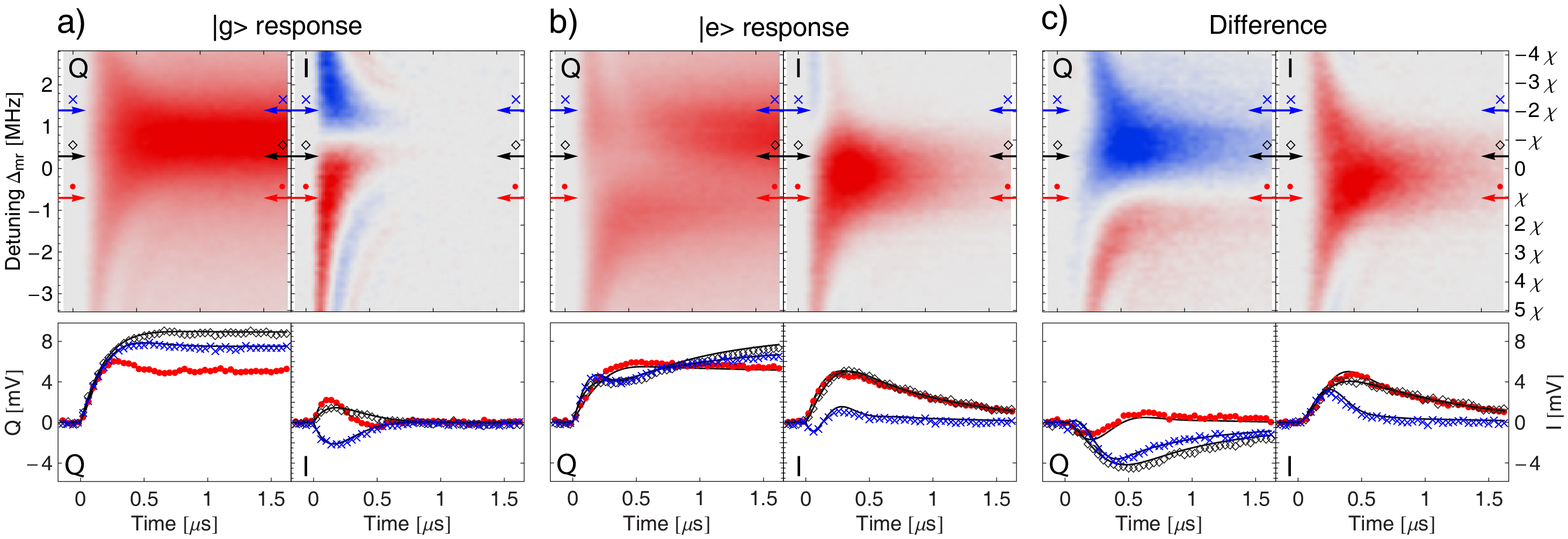}
    \end{minipage}
    \caption{(color online) $I$ and $Q$ quadratures for pulsed measurements at different detunings from the resonance frequency. The plots in a) are taken with the qubit in the ground state $| g \rangle$ while b) displays the response of the system with the qubit prepared in the excited state $| e \rangle$. c) shows the result of the pointwise difference of the data acquired with the qubit in the ground state and the excited one. The lower panels show time traces taken at different detunings (blue crosses: $\Delta_{{\rm mr}} = 1.4~{\rm MHz}$, black diamonds: $\Delta_{{\rm mr}} = 0.3~{\rm MHz}$, red dots: $\Delta_{{\rm mr}} = -0.7~{\rm MHz}$) with comparison to theory (solid lines).}
  \label{fig:pulsed_meas_detuning}
\end{figure*}

To avoid measurement-induced dephasing during the qubit manipulation most of the recent circuit QED experiments have been performed by probing the qubit state with pulsed measurements~\cite{Majer2007, Schreier2008, Chow2009, Leek2009, Filipp2009b, DiCarlo2009}. In contrast to a continuous measurement, the measurement tone is switched on only after the qubit state preparation is completed, see Fig.~\ref{fig:pulsed_iqtrace}a) for the pulse scheme.  The absence of measurement photons during qubit manipulation also avoids the unwanted ac-Stark shift of the qubit transition frequency, thus simplifying qubit control.

With the qubit in $|g\rangle$ the resonator response reaches its steady state at the rate $\kappa$, which is seen in the exponential rise of the $Q$ quadrature, see blue crosses in Fig.~\ref{fig:pulsed_iqtrace}b). Since the resonator is measured on resonance at its pulled frequency $\Delta_{{\rm mr}} = - \chi$, the $I$ quadrature is left unchanged, see blue crosses in Fig.~\ref{fig:pulsed_iqtrace}c). As in the continuous case, the resonator frequency is pulled to $\omega_{{\rm r}}+\chi$ when the qubit is prepared in $\ket{e}$, see red dots in Figs.~\ref{fig:pulsed_iqtrace}b) and c). Since the resonator is now effectively driven off-resonantly, the transmitted signal has non vanishing $I$ and $Q$ quadrature components both of which contain information about the qubit state. With the measurement frequency still at $\Delta_{{\rm mr}} = - \chi$, ringing occurs at the difference frequency $(\omega_{{\rm r}} + \chi) - \omega_{{\rm m}}= 2\chi$. At later times, the average response is approaching again the steady-state value as the qubit decays to $|g\rangle$ at the rate $\gamma_1$. As in the continuous case, the qubit lifetime in presence of measurement photons is obtained from a fit to the Cavity-Bloch equations. Note, that the decay of the quadrature amplitudes shown in Figs.~\ref{fig:pulsed_iqtrace}b) and c) does not directly correspond to the exponential decay of the qubit population $\langle \hat{\sigma}_{{\rm z}}\rangle$, but rather is determined by the interplay of resonator and qubit evolution.

The dynamics of the $I$ and $Q$ quadrature amplitudes can also be represented in a phase-space plot, see Fig.~\ref{fig:pulsed_iqtrace}d). The response for the qubit in $|g\rangle$ follows a straight line while the response for the qubit in $|e\rangle$ is more complex. The nontrivial shape of this curve reinforces the fact that both field quadratures contain information about the qubit state. It is obvious that a simple rotation in the $IQ$-plane cannot map the signal into a single quadrature.

Data taken at different measurement frequencies are shown in Fig.~\ref{fig:pulsed_meas_detuning}. As in Sec.~\ref{Sec:ContMeas}, the $I$ and $Q$ components are rotated such that $Q=A$ and $I=0$ in steady-state. For the theoretical curves (solid lines), the same set of parameters as for the analysis of the continuous measurement are used, leading to very good agreement. Fig.~\ref{fig:pulsed_meas_detuning}a) shows the measured response for the qubit in $|g\rangle$. The $Q$ quadrature shows the expected exponential rise in the cavity population and for $t \gtrsim 0.5~\mu$s we recover the continuous measurement response. The $I$ quadrature shows the transient part of the response during the initial population of the resonator, having a negative value (blue crosses) for measurements at a frequency above $\omega_{{\rm r}}-\chi$ (blue detuned) and a positive value (red dots) at frequencies below $\omega_{{\rm r}}-\chi$ (red detuned). Ringing can be observed when the measurement is off-resonant from the pulled cavity frequency. Fig.~\ref{fig:pulsed_meas_detuning}b) shows the response with the qubit prepared in $| e \rangle$. The response is similar to the one shown in Fig.~\ref{fig:cont_meas_detuning} for the continuous measurement, if one omits the initial~100~ns where the resonator is populated.

\section{Reconstruction of qubit state} \label{Sec:State_Reconstruction}

The detailed understanding of the dynamics of the dispersively coupled qubit/resonator system can be used to infer the qubit excited state population $p_e = (\langle \hat{\sigma}_{{\rm z}} \rangle + 1 ) / 2$. Indeed, the difference in the measured response for a given unknown state $s_{\rho}(t)$ and the ground state response $s_{g}(t)$, which corresponds to the shaded area indicated in Fig.~\ref{fig:pulsed_iqtrace}b) and c), is directly proportional to $p_e$.

To explicitly state this relation, we introduce an effective qubit measurement operator $\op{M}^{i}(t)$, an approach that we have already employed to perform two-qubit state tomography using a joint dispersive read-out~\cite{Filipp2009b}. Here, $i=I,Q$ denote the $I$ and $Q$ field quadratures used to measure the qubit state. In terms of this measurement operator, the $I$ and $Q$ components of the signal $s_{\rho}^i(t)$ for the qubit in state $\rho$ before the measurement are  given by
\begin{equation}
\label{eq:signal} s_{\rho}^{i}(t) \equiv \langle \op{M}^{i}(t)\rangle = {\rm Tr}[\rho \op{M}^{i}(t)],
\end{equation}
where $\op{M}^{i}(t)$ is determined by the solution to the master equation (\ref{eq:Master}). Analytical solutions can be found in the limit of vanishing qubit decay~\cite{Filipp2009b},
\begin{subequations} \label{eq:Mtop}
 \begin{eqnarray}
 \op{M}^{I}(t) & = & \epsilon \frac{ e^{-\kappa t/2}\left [2 \op{\chi} \cos \left (\op{\chi} t \right )+ \kappa \sin \left (\op{\chi} t \right )\right]-2 \op{\chi}}{\op{\chi}^2 + (\kappa/2)^2}, \\
 \op{M}^{Q}(t) & = & \epsilon \frac{ e^{-\kappa t/2}\left[\kappa \cos \left (\op{\chi} t \right ) - 2\op{\chi}\sin \left (\op{\chi} t \right )\right]-\kappa}{\op{\chi}^2 + (\kappa/2)^2},
 \end{eqnarray}
\end{subequations}
which depend on the operator  $\op{\chi} \equiv \Delta_{{\rm rm}} + \chi \op{\sigma}_{{\rm z}}$ for the qubit state-dependent cavity pull. As a consequence of performing a quantum non-demolition measurement with only a few photons populating the resonator, mixing transitions between the two qubit states can be neglected~\cite{Blais2004} and $\op{M}^{i}(t)$ is diagonal at all times. The qubit then remains in an eigenstate during the measurement~\cite{Boissonneault2009} and we can write $\op{M}^{i}(t) = s_g^{i}(t) \ket{g}\bra{g} + s_e^{i}(t) \ket{e}\bra{e}$. The signals $s_{g}^{i}(t) = {\rm Tr}[\ket{g}\bra{g} \op{M}^{i}(t)]$ and $s_e^{i}(t) ={\rm Tr}[\ket{e}\bra{e} \op{M}^{i}(t)]$ are determined by Eq.~\eq{eq:Mtop} for the values $\langle\op{\chi}\rangle = \Delta_{{\rm rm}} \pm \chi$ corresponding to the qubit in the ground or excited state. To account for qubit relaxation, the Cavity-Bloch equations (\ref{eq:cavity_bloch_n}) are solved to determine $s_{g/e}^{i}(t)$.

The qubit excited state population $p_e(\rho)$ in a given state $\rho$ is determined by the normalized area between the measured signal $s_\rho^{i}$ and theoretical ground state response $s_g^{i}$,
\begin{equation} \label{eq:area}
    p_{e}(\rho) = \frac{1}{T}\sum_j \frac{s_{\rho}^{i}(t_j) - s_g^{i}(t_j)}{s_e^{i}(t_j) -s_g^{i}(t_j)}  \Delta t,
\end{equation}
where $\Delta t$ denotes the discrete time step between datapoints. $s_{g/e}^{i}(t_j)$ are solutions to the Cavity-Bloch equations (\ref{eq:cavity_bloch_n}) with independently determined parameters. Replacing $s_\rho^{i} (t_j)$ with the corresponding expressions from Eq.~\eq{eq:signal}, we notice that Eq.~\eq{eq:area} simplifies to $p_{e}(\rho) = {\rm Tr}[\rho \ket{e}\bra{e}]$, demonstrating that the excited state population of an arbitrary input state is proportional to the normalized area between signal and ground state. Thus, the effective measurement operator $\op{M}^{\prime\,i} = \ket{e}\bra{e}$ defined by this procedure is equivalent to a projective measurement of the excited qubit state.

The measurement protocol can be summarized as follows: First, the relevant system parameters are determined in separate measurements as discussed in Section \ref{Sec:ContMeas}. The qubit lifetime $T_1$, the single remaining parameter, is determined by applying a $\pi$-pulse to the qubit and analyzing the resulting transmitted signal. From this complete set of parameters, the signals $s^i_g(t)$ and $s^i_e(t)$ are computed. Finally, the excited state population $p_e$ is calculated from the recorded signal $s^i_\rho(t)$ of an arbitrary qubit state $\rho$ and the theory lines $s^i_{g}(t)$ and  $s^i_{e}(t)$, using Eq.~\eq{eq:area}, which amounts to a measurement of $\op{M}^{\prime\,i} = \ket{e}\bra{e}$. For the particular case of the qubit being in $\ket{e}$ after a $\pi$-pulse, the point-by-point difference signal is shown in Fig.~\ref{fig:pulsed_meas_detuning}c). Note, that the excited state population can also be directly inferred from a fit of the Cavity-Bloch equations to $s_\rho^{i}(t)$ with $p_e$ as free fit parameter. It is, however, computationally less intensive to calculate the population with the area method from Eq.~(\ref{eq:area}), that is, to perform algebraic operations for the data analysis rather than employing a non-linear fit-routine. We have checked that both techniques provide the same results within the experimental precision.

\begin{figure}[b!]
  \centering
    \begin{minipage}{0.8\linewidth}
        \includegraphics[width=1\textwidth]{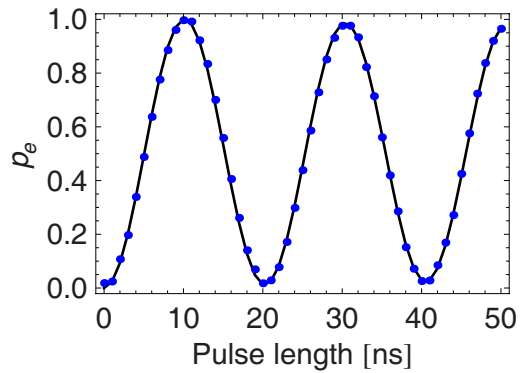}
    \end{minipage}
  \caption{(color online) Rabi oscillations of the qubit population $p_e$ reconstructed by analyzing with Eq.~\eq{eq:area} the pulsed $Q$ response of the resonator (dots). The black line corresponds to the theoretical prediction calculated using Cavity-Bloch equations with the following parameters: $\omega_{{\rm a}} / 2 \pi = 4.504~{\rm GHz}$, $\chi / 2 \pi = -1.02~{\rm MHz}$, $T_1=860~{\rm ns}$, $\Omega / 2 \pi = 50~{\rm MHz}$.}
    \label{fig:pop_mapping}
\end{figure}

To test our method experimentally, we perform a Rabi-oscillation experiment~\cite{Wallraff2005}, where a pulse of variable length $\tau$ and amplitude $\Omega$ is applied at the effective qubit transition frequency $\omega_{{\rm s,res}}$. Indeed, the population $p_e$ obtained with the area method (Fig.~\ref{fig:pop_mapping}, points) has an rms deviation of less than $1\%$ from the population predicted by Eq.~(\ref{eq:rabi}) (Fig.~\ref{fig:pop_mapping}, solid line). The data is also in good agreement with a simplified expression
\begin{equation} \label{eq:rabi}
p_e (t) \cong \frac{1}{2} - \frac{1}{2} e^{-\frac{t}{4} (3 \gamma_1 +
  2 \gamma_\phi)} \cos (\Omega t/2).
\end{equation}
predicting the time-dependent population of the qubit in the limit of large driving fields ($\Omega \gg \gamma_1, \gamma_{\phi}$)~\cite{Allen1987}.

\section{Conclusion}

In conclusion, we have presented a simple set of equations describing the dynamics of the average values of the quadrature amplitudes of the transmitted microwave fields in dependence on the qubit state in a circuit QED setup operated in the dispersive regime. The measured time dependent response of the cavity field to a change in the qubit state is in excellent agreement with calculations. The dependence of the measured response on measurement frequency is well understood both for continuous and pulsed measurements. From the time dependent measurement response we reconstruct the qubit excited state population that is used in tomographic measurements to accurately measure both single and two-qubit density matrices~\cite{Filipp2009b, Leek2009}.

\section*{Acknowledgments}

We acknowledge useful discussions with Maxime Boissonneault. We also acknowledge the group of M.~Siegel at the University of Karlsruhe for the preparation of Niobium films. This work was supported by the SNF project number 111899 and ETH Zurich. A.B. was supported by NSERC, CIFAR, and the Alfred P. Sloan Foundation.

\bibliographystyle{apsrev}
\bibliography{references}
\end{document}